# A Network Science Approach to Driver Gene Detection In Human Regulatory Network Using Gene's Influence Evaluation

Mostafa Akhavan Safar [a], Babak Teimourpour [a, *], Mehrdad Kargari [a]

[a] Department of Information Technology Engineering, Faculty of Industrial and Systems, Tarbiat Modares University, Tehran, Iran

Date :22 january 2020

## Abstract

Cancer disease occurs because of a disorder in the cellular regulatory mechanism, Which causes cellular malformation. The genes that start the malformation are called Cancer driver genes (CDGs) . Numerous computational methods have been introduced to identify cancer driver genes that use the concept of mutation.Regarding abnormalities spread in human cell and tumor development, CDGs are likely to be the potential types of gene with high influence in the network. This increases the importance of influence diffusion concept for the identification of CDGs.recently developed a method based on influence maximization for identifying cancer driver genes. One of the challenges in these types of networks is to find the power of regulatory interaction between edges.The current study developed a technique to identify cancer driver gene and predict the impact of regulatory interactions in a transcriptional regulatory network. This technique utilizes the concept of influence diffusion and optimizes the Hyperlink-Induced Topic Search algorithm based on the influence diffusion. The results suggest the better performance of our proposed technique than the other computational and network-based approaches.

## 1-Introduction

### 1.2. Driver gene detection problem

Cancer refers to any one of a large number of diseases characterized by the development of abnormal cells [1]. Abnormal cell growth and production help cause cell cancer. Cancer often has the ability to spread throughout your body.Although the mechanism has not yet been clarified, but it was found the involvement of genetic and other factors in the cell disturbance. According to

* Corresponding author. Room No. 210, 2nd Floor, School of Systems and Industrial Engineering, Tarbiat Modares University (TMU), Chamran/Al-e-Ahmad Highways. Intersection, Tehran, P.O. Box 14115-111, Iran.
*E-mail address:* b.teimourpour@modares.ac.ir (B. Teimourpour)

World Health Organization (WHO), cell cancer is the second-leading cause of death, accounting for almost 6.9 million deaths of 2018 in which one died out of six [2]. Lung cancer (2.09 million cases), breast cancer (2.09 million cases), and colorectal cancer (1.8 million cases) were the most common types of cancers diagnosed in 2018 [2]. Since there is no evidence confirm causes of cancer and the increasing rate of mortality, investigation into cancer has long been the focus of many researchers. The experiments in this area focused in particular on cancer drive gene (CDG) identification that are the main factors responsible for the cause of cancer in human cells. The purpose of finding these genes is to develop new drugs as a crucial target for main genetic influences on cancer disease, and to prevent the cell disturbance. The experimental methods for the identification of CDG suggests the extraction of different tumor cells of samples to apply a sequencing method for them. Significant mutated genes in most samples would be, thus, the CDGs. Because of their high costs and time-consuming processes, various computational methods have been developed. For the majority of these methods based on the analysis of transcriptomic or genomic data, the aim is to identify those genes with essential alterations in the gene expression. Simon and Youn (2011) introduced the Simon method [3]. The method was employed to improve identification of CDGs by estimating the background mutation rate and to predict altered background mutation rate among tumors and the redundancy of genetic codes of cancer genes with the analysis of mutations functional impact on proteins. One of the advantages of Simon method was able to distinguish between those mutations that affect protein function and those that do not, and recognize the difference between samples background mutation rate and diseased background mutation rate. The Oncodrive-FM (OFM) approach was proposed by Gonzalez-Perez and Lopez-Bigas (2012) [4]. Among different types of mutations, identifying CDGs and the pathways is an important challenge exists in cancer genomic area. Due to the limitations imposed by the traditional methods including difficulties of estimating the accurate mutation rate and growth-dependent alterations, a new criterion, FM bias, was developed in OFM approach which does not depend on the recurrence and can successfully identify CDGs. A novel approach, MeMo, was presented by Ciriello et al., (2012) to systematically analyze oncogenic pathway module. The method used correlational analysis, statistical tests, and three criterions were used for identifying the network modules. The three criterions were including those genes altered in the tumor samples, all tumor samples gene that involved in an identical biological process, and those genes that altered exclusively. Another approach was introduced by Zhao et al., (2012) . This technique, MDP finder, tries to solve the maximum weighted matching problem designed for identifying mutated driver pathway (MDP)[5]. This procedure involved genetic-algorithm approach and an integrated method consists of the combined gene expression and mutation data to find MDPs and to identify CDGs. In the study by Bashashati et al., (2012), a new technique, DriverNet, was introduced . Bashashati et al., established a computational framework in this method for the identification of driver mutations through microRNA (miRNA) expression networks. In DriverNet approach, the relationship between genome aberrations and transcription patterns were extracted through the gene interaction. Dendrix method was presented by Vandin et al., (2012) [6] to distinguish driver from passenger mutations. Dendrix method was combined with coverage and exclusivity which refers to a technique of gene identification in different samples and uncommon mutations in particular samples, respectively. Reimand et al., (2013) suggested ActiveDriver method . Based on this method, most CDGs were identified by their frequent mutations in different tumor samples. There are, however, some mutations that reveal an important functional role in the proteins. Exploring these mutation and functional protein locations were performed to enhance the identification of CDGs and predict the mutations mechanism. An unsupervised approach, iPAC, was developed by Aure et al., (2013) .The described systematic approach was based on an integrative analysis of gene expression data and number of copies to use a sequence of the statistical test on a selected gene list and extract matched CDGs out of the list. Lawrence et al., (2013) developed a method that involved a solution procedure for inhomogeneity problem of mutation process and frequency to discover gene abnormalities and, thus, identify CDGs. The identification of CDGs was facilitated by assessing transcriptional activities of the gene, comparing the observed frequency of mutations in types of cancer and human genome. The method of OncodriveClust was proposed by Tamborero et al., (2013), In this method, those genes with significant bias toward clustering of mutations in the sequence of protein were identified. Besides, a classification model for genes were built using coding-silent mutations measurement. Porta-Pardo and Godzik (2014) developed the technique of e-Driver . Based on this method, the internal dispersion of malignant mutations among functional regions in proteins were extracted to calculate the mutation rate in these regions compared to the others. If these investigated regions demonstrated positive results, then those genes could be CDGs. the DawnRank approach was suggested by Hou and Ma (2014) . It was a computational method to identify rare and specific gene types as CDGs using the mutations data and investigating all samples carry cancer genes. In this method, personal medical information of only one sample was used to identify the cancer gene. The mutated genes were ranked according to their potential to be drivers in the molecular interaction network. Top-scoring mutated genes had likelihood of being identified as the drivers. Zhang et al., (2014) introduced a computational approach, CoMDP, in which those mutations operated through same pathways were studied for CDGs identification instead of focusing on the genes. These pathways that lead to the cancer genes should exhibit enough coverages. Besides, a long-range correlation should be observed in the pathways. Another method, called MSEA was introduced by Mäkinen et al., (2016) . The method involved a brief and integrated view of CDGs mechanism rather analyzing data set separately. For recovering the biological pathway and gene networks, the method was provided with a computational pipeline that integrates the multidimensional data with biological functions and molecular networks. The CDGs identification would be then carried out. Contrary to the previous computational and statistical approaches, Rahimi, Teimourpour et al., (2019) introduced iMaxDriver approach that used influence maximization concept for identifying CDGs . In addition, other approaches including GISTIC , MuSic , MutSig , CHASM , CONEXIC , PARADIGM-Shift, TieDIE , MAXDRIVER, VarWalker , MOGA were presented.

In this study, we compared our proposed method and the obtained results with the 20 methods that introduced in above. We then demonstrated the comparison results, indicating the better performance of our method. The investigated computational methods were including ActiveDriver, OncodriverCLUST, e-Driver, Oncodrive-fm, Simon, Dendrix, CoMDP, MDPFinder, MutSigCV, iPAC, MSEA that are based on transcriptome data analysis, mutation concept, and biological pathways for identifying CDGs. DriverNet, MeMo, NetBox, and DawnRank use biological networks in addition to the mutation concept. For identifying CDGs, the approach of iMaxDriver only embraces the biological networks concept. Although CDGs identification methods have been the subject of intense study, but they suffer limitations for the following reasons:

1) most of these proposed methods need mutation data. However, these types of data are in nature high-level noise.
2) the methods generally result in a high false positive rate and low precision and F-measure.
3) Most computational methods are costly and time consuming.
4) The majority of developed techniques fail to explore the advantages of certain essential information from the structure in biological networks and the related interactions of the networks by neglecting topological parameters of biological networks.
5) Although it uses only the network approach and the theory of diffusion maximization without using the concept of mutation,But due to the specificity of propagation algorithms, it is very time consuming and time consuming.Moreover, sort of impacts the regulatory interactions have will lead to different consequences in the transcriptional regulatory network. An accurate prediction is not fulfilled even by increasing weight of the edges randomly.

For these limitations, we proposed a technique of CDGs identification in human transcription regulatory network (TRN) that uses theoretical network-based diffusion without requiring the pathway data and mutation concept. Our technique performed better compared to the most of above mentioned techniques, particularly the iMaxDriver methods , in terms of time-consumption and obtained results. In addition, we introduced a method to predict impacts the regulatory interactions have in the network.

## 2-Authority and influence evaluation background

Analysis of information diffusion is an essential process in social networks [7]. Generally, diffusion implies whatever, physically or virtually, transmit from a node to another including epidemic diseases, gossip spread, product advertisement etc. Certain transmissions do not reveal the existence of whatever is transmitted such as transmitting diseases. In this case, only the group of infected individuals will be identified. For understanding the disease transmission, the information exchange is addressed through the network diffusion and available patterns. In other words, the information is transmitted through the network nodes. Various types of models have been designed for influence in the network by which the individual value or diffusion are examined in a social network. This is important for recognition of potential influencers [8,9]. Generally, when individual idea or activity impacts others', such diffusion transmission is called influence diffusion. Because diffusion in nature is an impact occur through output links, for instance persuading consumer to buy a product, the results' value or diffusion in individuals might appear differently. While value implies the number of input links for the followers, indicating the important value of the individual.

It has been demonstrated the degree of closeness ascribe to these both approaches, indicating individual gained value by affecting others [10]. There are different types of approaches to measure value and diffusion such as Hyperlink-Induced Topic Search (HITS) algorithm. This algorithm is a link analysis algorithm that can classify web pages. Jon Kleinborg was was the developer of HITS algorithm . In HITS algorithm, certain pages function as a central hub, for instance pages enriched in output links. Others are topic-based pages, means those with input links. This algorithm assigns each page two scores including hub and authority scores. Good hubs are linked to good authorities based on Kelinberg proposed model. We consider this algorithm consisting of a page $P$, a hub score ($h_p$), and an authority score ($a_p$) as follows:

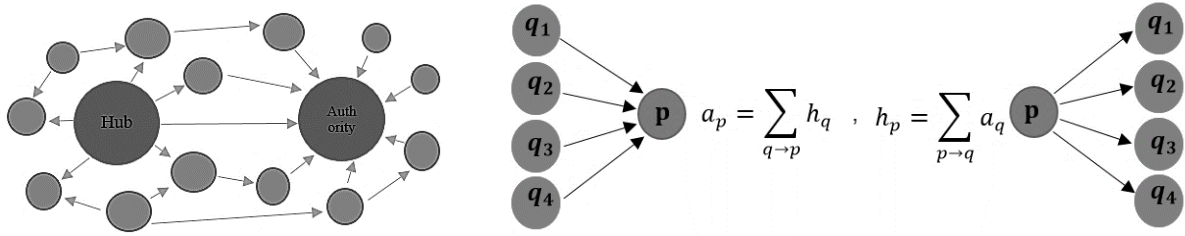

Figure 1: Calculation of Hub and Authority scores in HITS

$$a_p^{(k)} = \sum_{j:\, e_{qp} \in E} h_q^{(k-1)} \qquad (1)$$

$$h_p^{(k)} = \sum_{j:\, e_{pq} \in E} a_q^{(k)} \qquad for\ k = 1,2,3,\ldots \qquad (2)$$

Where $e_{pq}$ is a hyperlink from P page to q page, $E$ is the set of hyperlinks. From the above equations, each iteration of the algorithm includes:

Authority score update in which authority score of each node is assumed proportional to the sum of hub scores pointed at by the authority nodes. Hub score update in which hub score of each node is assumed proportional to the sum of authority scores pointed at by the hub nodes. For score of each page, assign 1 hub and authority scores in all pages. Algorithm will repeat the computation of hub scores and algorithm scores until convergence of the constant values. We have

$$\boldsymbol{a}^{(k)} = \boldsymbol{L}^T.\boldsymbol{h}^{(k-1)} = \boldsymbol{L}^T.\boldsymbol{L}.\boldsymbol{a}^{(k-1)} \qquad (3)$$

$$\boldsymbol{h}^{(k)} = \boldsymbol{L}.\boldsymbol{a}^{(k)} = \boldsymbol{L}.\boldsymbol{L}^T.\boldsymbol{h}^{(k-1)} \qquad (4)$$

Here, $\boldsymbol{a}$ and $\boldsymbol{h}$ denote column vectors identifying the authority and hub scores, $\boldsymbol{L}$ is adjacency matrix for the system adjacency, so that

$$L_{ij} = \begin{cases} 1 & if\ (i,j) \in E \\ 0 & otherwise \end{cases}$$

In weighted network, the formula is modified as follows:

$$a_p^{(k)} = \sum_{q:\, e_{qp} \in E} w_{qp} h_q^{(k-1)} \quad (5)$$

$$h_p^{(k)} = \sum_{q:\, e_{pq} \in E} w_{pq} a_q^{(k)} \quad for\ k = 1,2,3,\dots \quad (6)$$

$$\boldsymbol{a}^{(k)} = \boldsymbol{L}^T.\boldsymbol{W}^T.\boldsymbol{h}^{(k-1)} = \boldsymbol{L}^T.\boldsymbol{W}^T.\boldsymbol{L}.\boldsymbol{W}\boldsymbol{a}^{(k-1)} \quad (7)$$

$$\boldsymbol{h}^{(k)} = \boldsymbol{L}.\boldsymbol{W}.\boldsymbol{a}^{(k)} = \boldsymbol{L}.\boldsymbol{W}.\boldsymbol{L}^T.\boldsymbol{W}^T.\boldsymbol{h}^{(k-1)} \quad (8)$$

## 3-The study network and methodology

### 3.1. Transcriptional regulatory network (TRN)

As was mentioned already, one reason for cancer incidence is disturbance in regulatory intramolecular interactions. Accordingly, studying the relations which can be taught of as a biological process can lead to the identification of disturbance factors and can provide a systematic perspective rather molecular. Transcriptional regulatory network is one of the most challenging type of regulatory biological network in which a functional impairment can lead to cancer incidence. Gene regulatory network is a type of biological network which is interpreted by gene expression. In this network, transcription factors impact other genes. Transcription factors (TFs) are the most important contributors to gene regulation and are prominent components in cells that control gene expression. TFs help determine the pattern to assume cells function and their respond to environment. The impaired function of regulatory TFs would raise the possibility of disease spread, particularly cancer incidence [11]. Analysis of this network and the relationships among TF-Target are helpful for understanding the information of a biological system and complex characteristics such as human disease. This network is a directed type of network for which the nodes, genes, TFs, and the edges appear in the physical or regulatory interactions among nodes. This kind of networks are required for identifying cancer key factors [ 12, 13, 14, 15].

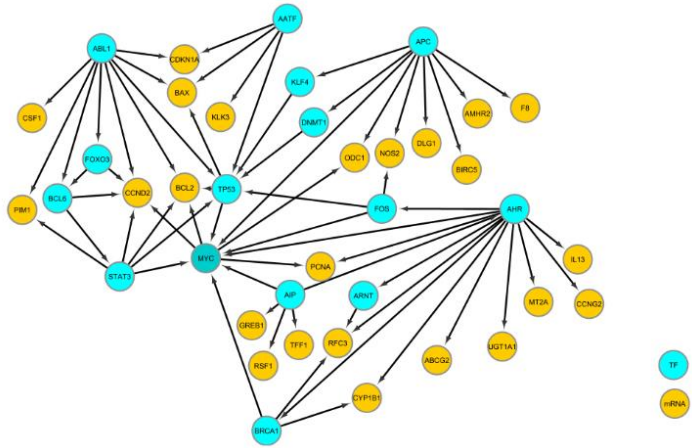

**Figure 2**: An overview of the TRN network structure in this study

We proceeded the network modeling with the online relationship database of TRUSST v2 [16] in this study. By using mining approach for extracting data, the TRUSST network v2 provides 9396 regulatory relationships of 2862 genes and TFs, including 795 TFs and 2067 genes. Neither weight nor interaction coefficient have been reported for TRUSST network relationships. Figure 3 illustrates a schematic view of this network.

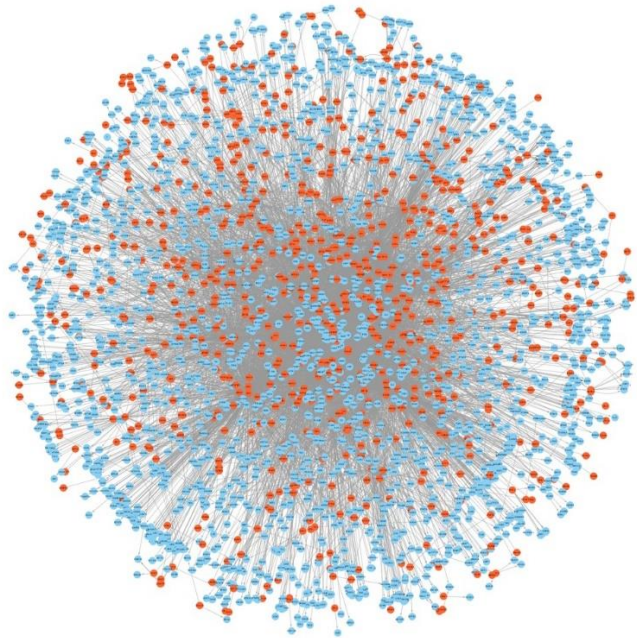

Figure3: TRRUST Network[ https://www.grnpedia.org/trrust/]

*3.2 Research methodology*

We noted earlier that deficit in the regulatory function of human cells could be caused by gene abnormalities and their impact on the others. Hence, a perspective of network influence diffusion would imply a work on this subject. Regarding this and the structure of TRN, those genes that appear to highly influence the others are potentially CDGs. It seems possible to identify CDGs and generalize the influence diffusion concept through the regulation of HITS algorithm based on influence diffusion concept in gene regulatory networks. Applying diffusion-based algorithm on gene regulatory networks are reasonable due to high sparse features of these networks. In this case, hub scores achieved on each gene in which higher score of a gene justify influence diffusion concept of others. It is because of TFs which reveal gene abnormalities in the transcription regulatory network and impact themselves and mRNA. Besides, TFs share structural similarities with the gene regulatory network. In such network, the genes with highest influence on the others are those that are involved in spread of gene abnormalities and cancer incidence.

To our knowledge, no investigation has employed the approach of Hyperlink-induced topic search for identifying CDGs in TRNs. To utilize this approach in biological networks, we applied biological characteristics of gene in algorithm. This enabled the algorithm to consider the biological principles of the genes regulatory behavior in network. We assigned each of gene scores based on the first authority and hub scores. In the original version of HITS algorithm, hub and authority are assigned scores equal to 1; while in our investigation, both scores were calculated based on gene expression rate in normal and cancer tissues which will be discussed later. The impacts of regulatory interactions in TRNs has largely been ignored. Number of investigations such as iMaxDriver method examined TRNs with random edge weights. However, studies with random edge weights might not provide an accurate impacts of regulatory interactions in networks. In our study, we calculated the impacts of regulatory interactions through the application of load-balancing theory of the network. During calculation of the influence and diffusion of each gene, we applied the obtained impacts in the algorithm. Hence, the concept of influence diffusion in calculation, initial scores, and application was applied based on biological characteristics. Each hub score was calculated and soreted in descending order after running the algorithm and constructing the network. The highest ranking gene were then compared to the standard cancer data set, and ultimately compared to other computational methods proposed for the identification of CDGs. The results showed that our proposed technique had better performance than computational and diffusion-based methods.

## 4- Constructing weighted interaction network

A list of regulatory interactions was downloaded from TRUSST[16] data set for constructing a TRN. To construct cancer-type specific network and to calculate initial genetic scores of the algorithm as well as each regulatory interaction weight, the required gene expression data was downloaded from the GEO database. To compare our method with those that have been proposed recently, cancer gene expression data including breast cancer (GSE15852), lung cancer (GSE232323), and colorectal cancer (GSE3268) was also downloaded. After analyzing the data, three data files were obtained, each representing a specific type of cancer. Each file contained a gene name and its expression rate data for normal and cancer tissues in a group of five patients. Networks constructed for the three types of cancer based on the regulatory interactions obtained from the TRUSST along with the three data files mentioned above. For constructing the initial cancer network, an interaction without expression value of the source and target gene in the related file was filtered and the other remained. We used the expression rates to assign weight to each gene in the constructed networks. Means the difference between the rate of expression in a normal and cancer tissue of a patient was considered the gene weight.

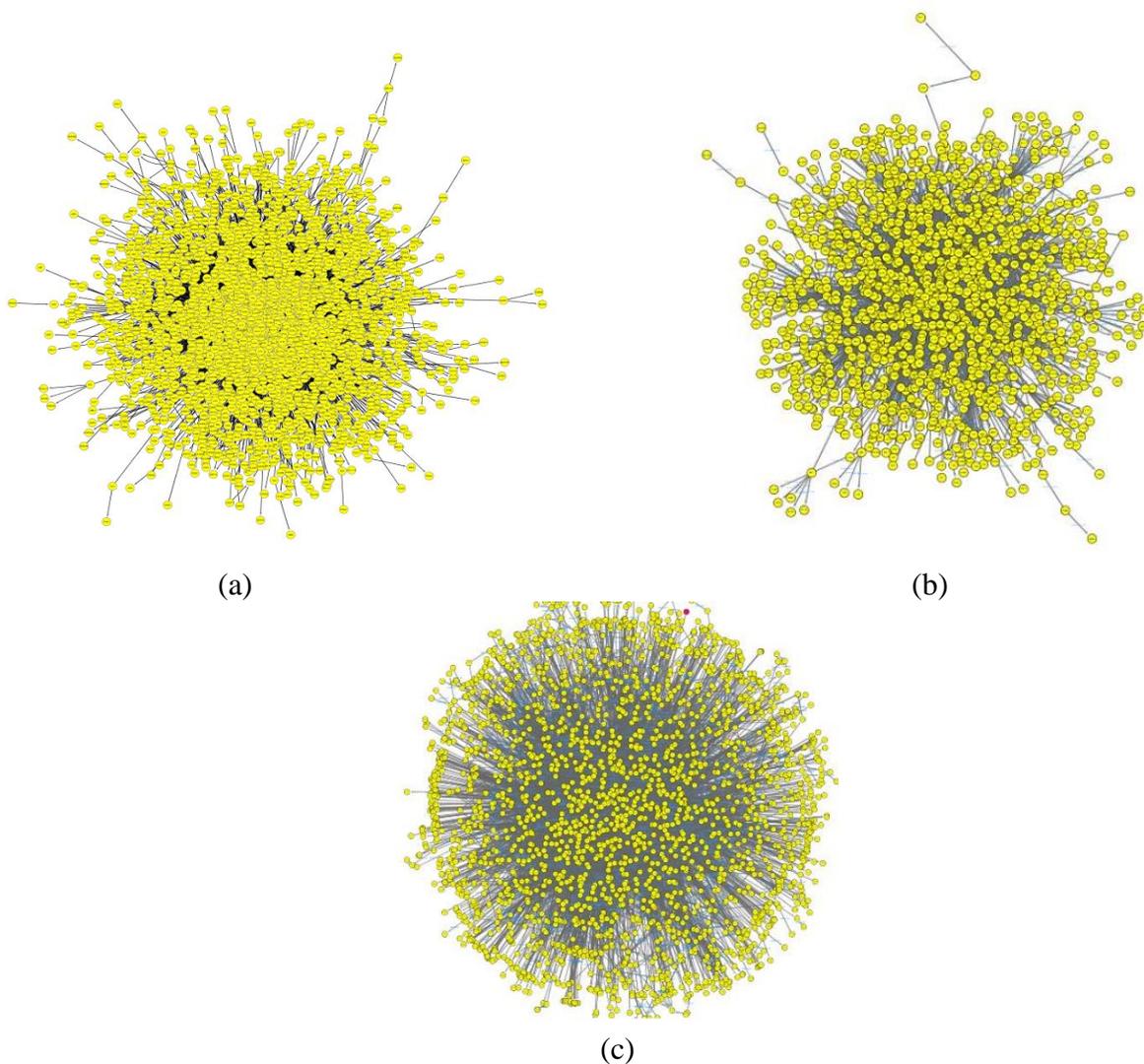

(a) (b)

(c)

Figure 4: The cancer networks to study after preparing and filtering :(a)LUSC Network ,(b) COAD Network , (c) BRCA Network

The purpose of this experiment was to test whether gene expression alterations impact expression behavior of others. Means a high different expression level of a gene would spread a degree of influence on its neighbors. For constructing a weighted interaction network, we used the theory of load-balancing based on the network structure. This allowed us to assign weight to each interaction based on the difference of expression, indicating gene natural behavior in the network. Based on this theory, we assumed that the target gene alterations are equal to those of source gene to the extend the source gene influences the target gene. Hence, the influence diffusion concept can be considered in the weighting from a biological perspective. Since gene weights were assigned before, vector $\vec{v}$ was formed which consists of all network weights. The matrix of weight, $L$, was then formed. This matrix is the directed network adjacency matrix in which the entries for those genes with interactions are unknown. The proposed hypothesis in this study makes us compute the $Min||L^T.v - v||$. The solution to this equation would determine $L$ matrix unknown entries which were equal the extent to which source gene influenced the target gene. to obtain the solution of $Min||L^T.v - v||$, a reduction gradient approach was used as follow.

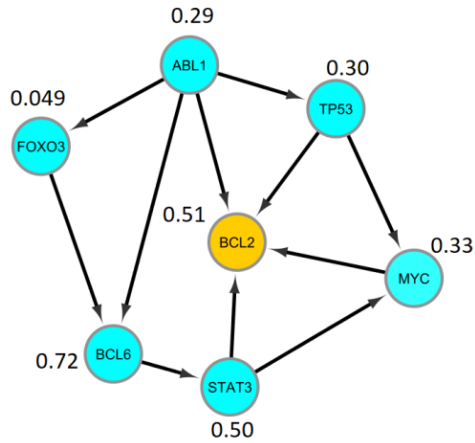

$$L = \begin{cases} w_{i \to j} & if\ (i,j) \in E \\ 0 & otherwise \end{cases}$$

$$V = \begin{bmatrix} w_{ABL1} \\ w_{FOX3} \\ w_{TP53} \\ w_{BCL2} \\ w_{BCL6} \\ w_{STAT3} \\ w_{MYC} \end{bmatrix} = \begin{bmatrix} 0.29 \\ 0.30 \\ 0.50 \\ 0.40 \\ 0.33 \\ 0.72 \\ 0.33 \end{bmatrix}$$

$L$ = weighted adjacency matrix

$\ell ost\_function = Min||L^T.v - v||$

$L^T.\vec{v} = \vec{v}$ is a matrix equation which assumed to have unknown entries. $\vec{v}$ is a column vector constitutes of constants and known values. We aim to find the unknown values of the matrix $L^T$ such that $L^T.\vec{v}$ is equal to the vector $\vec{v}$. We have no way of finding a deterministic solution to $L^T.\vec{v} - \vec{v} = 0$ since the determined weight for each gene was based on the calculation of gene expression changes. It simply reflects the fact that biological data in nature can cause this uncertainty. We therefore preferred a reduction gradient approach to reach an approximate solution. Hence we began by using a matrix $(L^T{}_0)$ with random elements for the unknown entries of matrix $L^T$. We then calculated $||L^T.\vec{v}_0 - \vec{v}|| = 0$ and used the redunction gradient method to calculate the next value $(L^T{}_1)$. The optimal solution for $||L^T.\vec{v}_0 - \vec{v}||$ is the answer where has the closest value to zero. Thus we put our focus to minimum the loss function.

Norm-2 of a vector is given by:

$$|V|_2 = \left( \sum_i |v_i|^2 \right)^{1/2} \quad (9)$$

Weights of the edges were derived from repeating the above procedure for lung, breast, and colorectal cancer networks. Initial scores were required for the ranking algorithm when the calculation of gene and interaction weights was completed. We assigned initial scores equal to the first rate of gene influence in the network, means the sum of weights of output edges from each node as:

$$h^{(1)}_{gene_i} = dexp(gene_i) \times \sum_{if:\, gene_i \to gene_j} w_{gene_i \to gene_j} \quad (10)$$

$$dexp(gene_i) = |Normal\_exp_{gene_i} - Canccer\_exp_{gene_i}|$$

Therefore, the proposed method consists of four main steps:
1) constructed the network using regulatory interaction database and gene expression data, after filtering and data preparing.
2) Gene weighting using genes using expression differences in normal and cancer tissue.
3) assigned interaction weight through minimizing the proposed influence-based lost Function.
4) initialization through influence concept .
5) Run the modified HITS algorithm and ranking the genes .

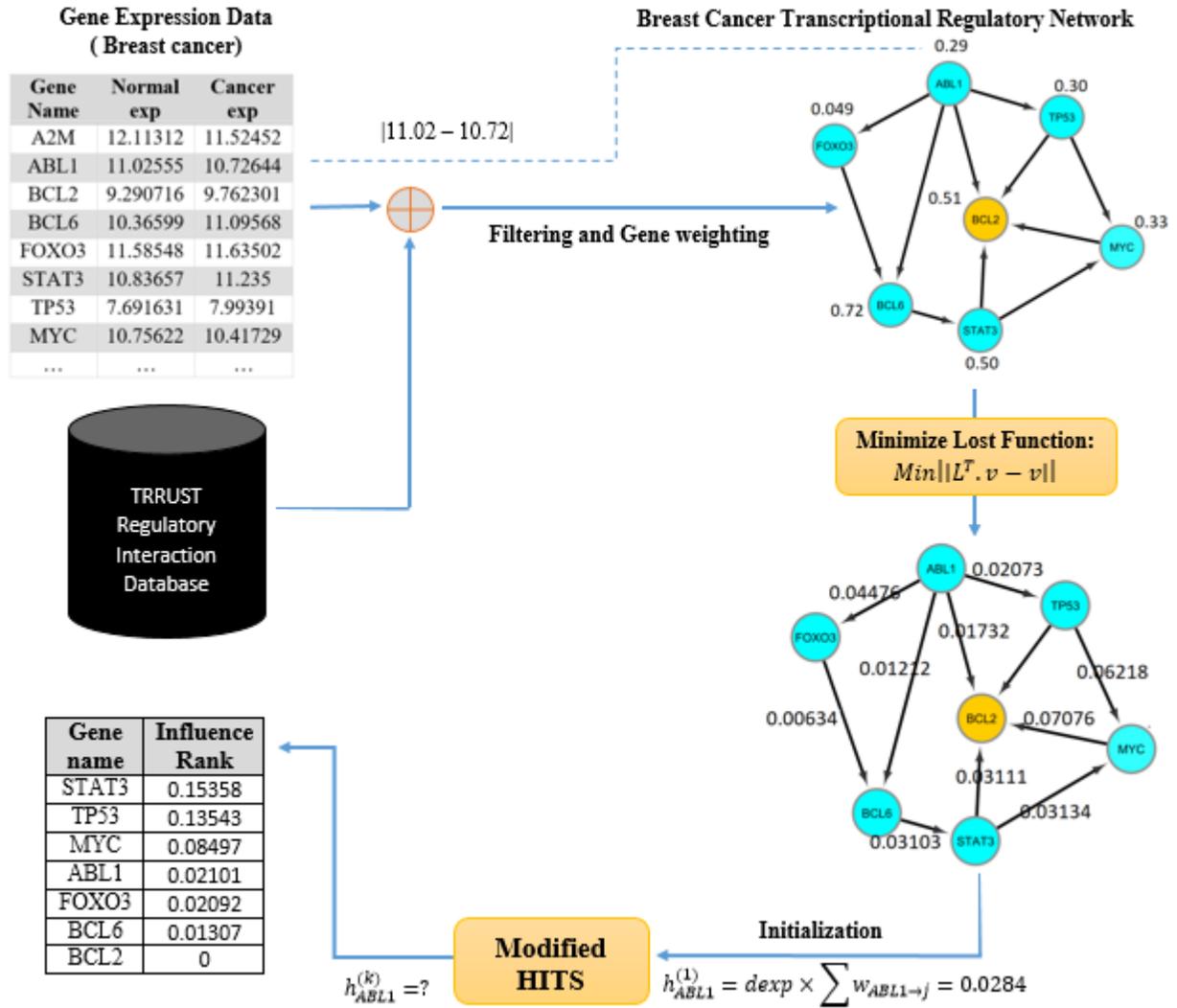

**Figure 5:** Overview of our proposed Method. 1) the network is constructed using TRRUST Regulatory Database and Gene expression Data. 2) weighting Genes using Expression differences in normal and tomoral tissue 3) weighting edges through minimizing the influence-based lost Function 4) initialization through influence concept 5) Run the modified HITS algorithm and ranking the genes.

## 5- Evaluation method

In this section, we compared our algorithm with the computational and network-based approaches to evaluate our method against the criteria mentioned in the previous sections. Every approaches in the comparison was run using DriverDB v2 [17], and the obtained results as well as those of our proposed technique were then compared with the standard data set of cancer gene TCGA [18] including breast (TCGA-BRCA), colorectal (TCGA-COAD), and lung (TCGA-

LUSC). The reported CDGs in this dataset were including 572,572, and 566 for breast, colorectal, and lung cancers, respectively.

The obtained results which were derived from existing criteria proposed in this study including the precision, recall, and f-measure criterion, for the evaluation of our algorithm and the other approaches are presented here. The use of recall criterion indicates the accuracy ratio of the identified CDGs to the total amount of reported CDGs. The precision criterion is for representing how precise the prediction of CDGs identification is. A good precision criterion can only be achieved if a high false-positive value is obtained. An important point here is the trade-off between these two criterion. Means an increased value of one of them will result in a decrease in the value of the other. Moreover, we will not be able to use a mathematical mean. Hence, a harmonic mean is the use of F-measure which reflects the accuracy of the experiment. F-measure considers both precision and recall criterion. The worst case value for F-measure is 0 and its best value is 1.

$$Recall = \frac{TP}{TP + FN} \qquad (11)$$

$$Precission = \frac{TP}{TP + FP} \qquad 12)$$

$$F - measure = 2 \times \frac{Precision \times Recall}{Precision + Recall} \qquad (13)$$

## 8- Results

In this study, a transcriptional regulatory network was evaluated for three types of cancer based on influence diffusion concept, gene expression data, and calculation of regulatory interaction impact. The obtained genes were then ranked using our proposed algorithm. The scores were sorted in descending order and And were separated based on a threshold value. They were classified as cancer causal and none cancer causal based on threshold value. Then We calculated Recall, precission and F-measure for each cancer network. Figure 6 shows the result in colorectal cancer network .our proposed method achieved the best F-measure compared to the other computational methods, indicating the improved performance of the prediction algorithm. This method even performed better that the latest suggested network based methods(iMaxDriver). The best results for the identification of CDGs was achieved by our proposed method by identifying 138 CDGs.

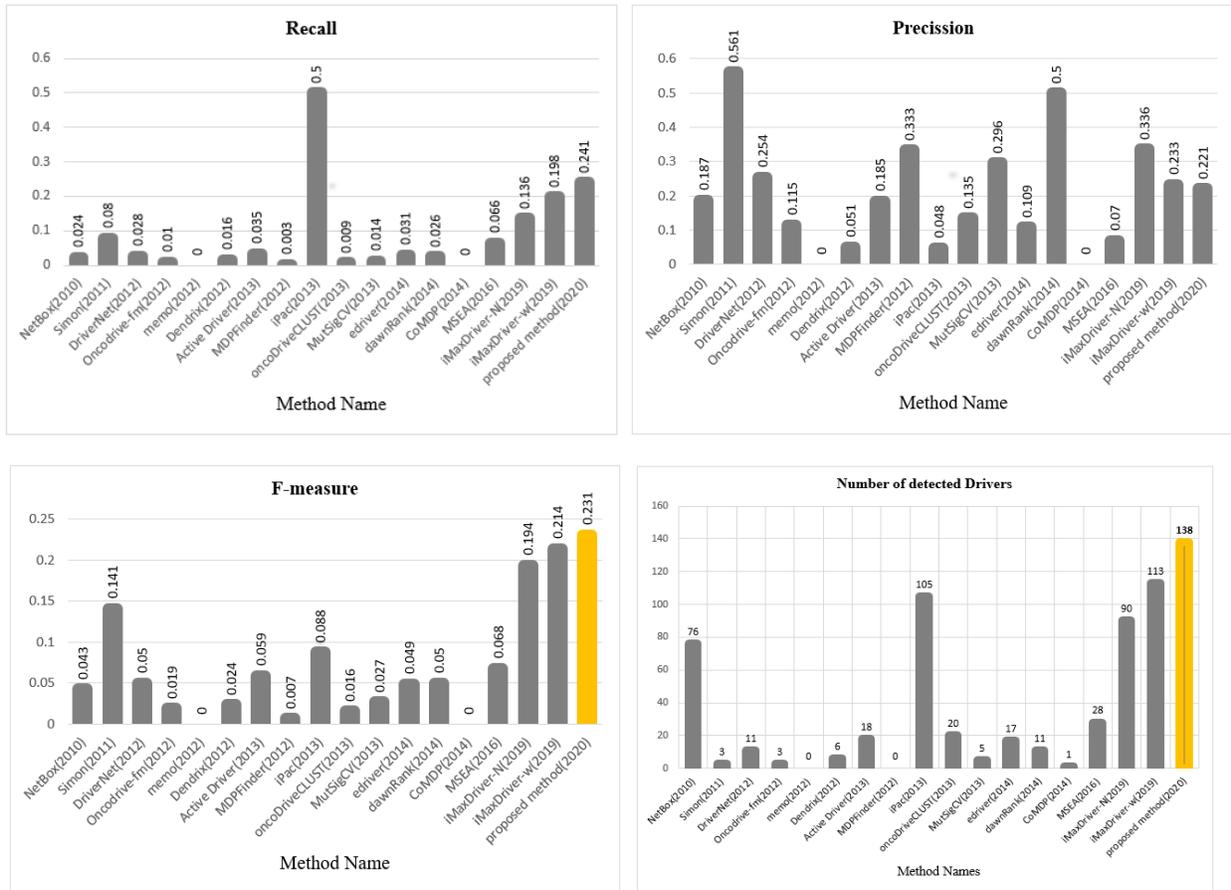

Figure 6: Comparison the results of proposed algorithm with other methods for COAD cancer.

In addition, in cancer breast network as show in Figure 7, our proposed method achieved the highest scores in F-measure compared to the other computational methods. This method even performed better that the latest suggested network based method(iMaxDriver). Also, compared with the iMaxDriver-N and iMaxDriver-w, The proposed method has higher F-measure and more detection power in terms of the number of CDGs, indicating improved performance of the prediction algorithm , and our proposed method achieved the best scores by identifying 132 CDGs among network based method and computational method after iPac.

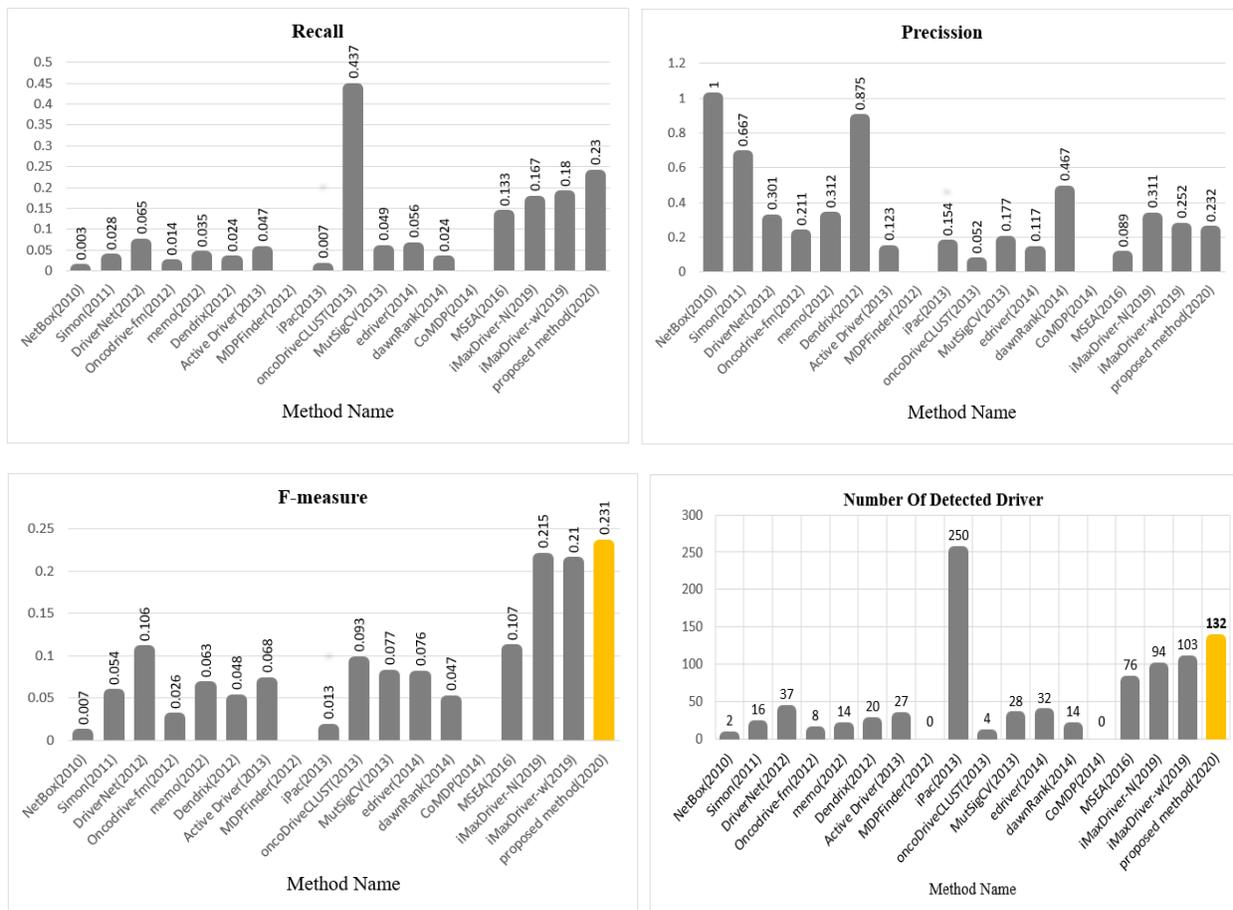

Figure 7: Comparison the results of proposed algorithm with other methods for BRCA cancer

In lung cancer network as shows in Figure 8 , the proposed method achieved a highest score in F-measure to the other computational methods. This method even performed better that the latest iMaxDriver, Which shows that the algorithm performs well in this network too.

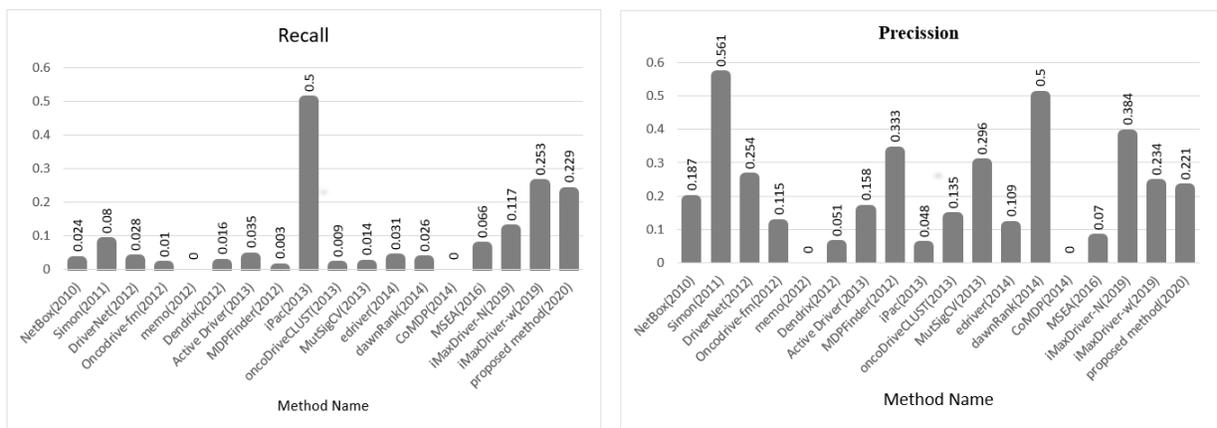

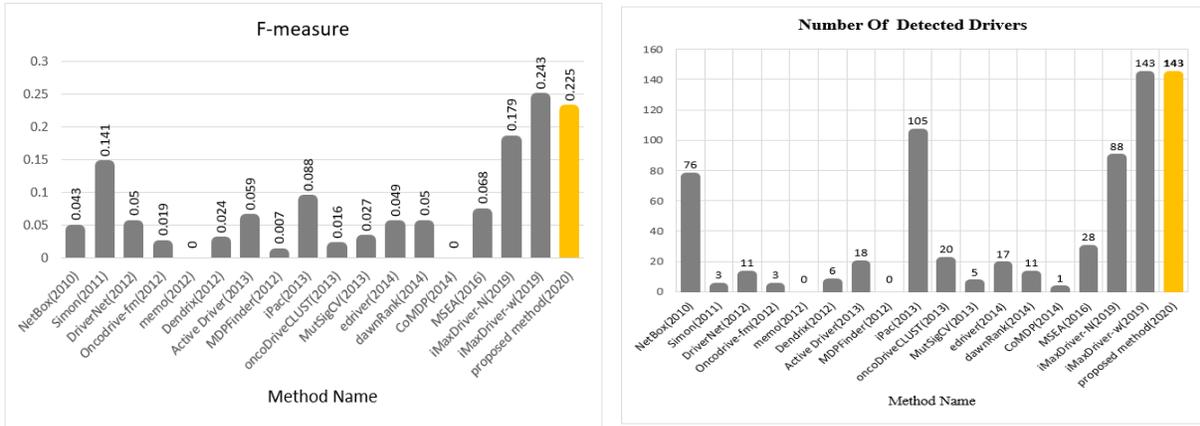

Figure 8: Comparison the results of proposed algorithm with other methods for LUSC cancer

In addition ,As shown in the veen diagrams. in Figure 9, our proposed method has detected new driver genes in all cancers. And so it works better than both the computational and the iMaxDriver networking methods. In addition, our proposed method has been able to cover a significant number of diagnostic genes by other methods.

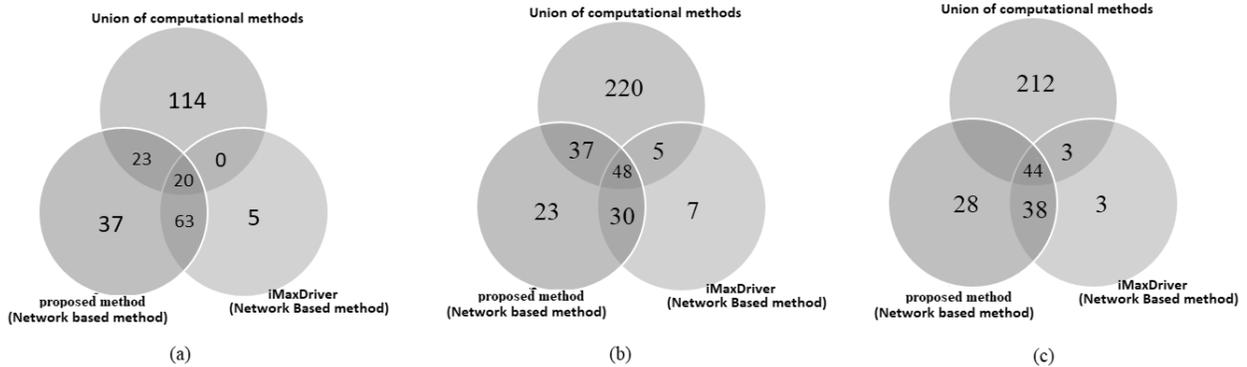

Figure 9:The Venn diagram of predicted CDGs using proposed algorithm and the union of computational methods and networking based method (a) LUSC ,(b) COAD ,(c) BRCA

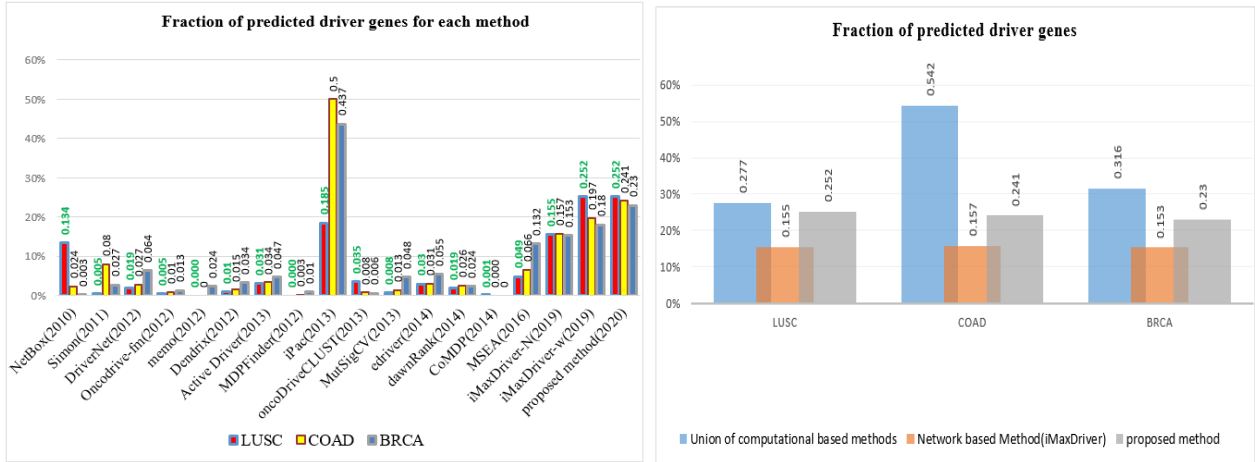

Figure 10 : Fraction of predicted driver genes for each method

## 9- Conclusion

In this study, we suggested a new algorithm for identifying cancer driver genes (CDGs) in a transcriptional regulatory network (TNF), using the influence diffusion concept and HITS weighted-based algorithm. Most efforts have used computational methods and mutation concept to identify CDGs. One of the problems in using these methods is the need for a huge amount of data. Another is the lack of accuracy in CDGs identification. It is the reason for using network-based approaches. A network-based approach is iMaxDriver that uses influence maximization. This approach however has some problems. Huge data consumption as well as complexity of the algorithm are two most important challenges in influence maximization. This approach, thus, is a time-consuming process. In addition to this, iMaxDriver approach was assigned with random regulatory interaction weights which might not provide an accurate prediction of the interaction impacts in a gene regulatory network. To address these limitations, our network-based proposed approach is developed in this study which requires low-data rate. We used the influence diffusion concept in a weighted-based HITS algorithm for developing this technique. To our knowledge, no effort has been done to identify CDGs of a TRN based on our proposed algorithm. We used biological characterestics of genes to apply the algorithm in a biological network. This enabled the algorithm to consider the biological principles of the genes regulatory behavior in network. Based on the first rate of gene influence and the gene expression rate in normal and cancer tissues, the initial scores of hub was calculated. Also, the influence diffusion was considered for the calculation of gene weights. Ultimately, the influence and diffusion for each gene was examined. We compared our proposed technique results with those of other techniques. According the F-measure scores and the identified CDGs, the proposed method achieved the best results in lung and colorectal cancers, compared to other developed approaches. For breast cancer, our approach is the second best method among the others. For all three types of cancers, our method performed better than iMaxDriver approach.